\def\ii{\'{\i}}

\def\beg{\begin{equation}}
\def\fim{\end{equation}}

\input epsf

\documentstyle[preprint,aps]{revtex}
\begin{document}

\title{Growth Exponent in the Domany-Kinzel Cellular Automaton}

\author{A.P.F. Atman and J.G. Moreira} 

\address{Departamento de F\ii sica, Instituto de Ci\^encias Exatas,\\
Universidade Federal de Minas Gerais, C. P. 702\\ 
30123-970, Belo Horizonte, MG - Brazil}
\date{\today}
\maketitle

\begin{abstract}
In a roughening process, the growth exponent $\beta$ describes how the roughness
$w$ grows with the time $t$: $w\sim t^{\beta}$. We determine the exponent
$\beta$ of a growth process generated by the spatiotemporal patterns of the
one dimensional Domany-Kinzel cellular automaton. The values obtained for 
$\beta$ shows a cusp at the frozen/active transition which permits determination
of the
transition line. The $\beta$ value at the transition depends on the scheme used:
symmetric ($\beta \sim 0.83$) or non-symmetric ($\beta \sim 0.61$). Using 
damage spreading ideas, we also determine the active/chaotic transition line;
this line depends on how the replicas are updated. 
\end{abstract}

\vskip 2truecm
\noindent{PACS:	05.10.-a, 02.50.-r}

\newpage

\section{Introduction}

The one dimensional Domany-Kinzel cellular automaton (DKCA) is a totally 
discrete sys\-tems - temporally, spatially and in numbers of states - 
with several applications in physics, chemistry, biology, computer
science, etc. The DKCA phase diagram was originally proposed  
by Domany and Kinzel \cite{dk} who showed the existence of two phases: 
an active phase and a frozen one. A more detailed study, using numerical 
simulation, was present by Martins {\it et al} \cite{martins},
where a new phase in the active region - a chaotic phase - was discovered using 
the damage spreading technique. Further, Zebende and Penna
\cite{zebende} used the gradient method to determine the phase boundaries 
with high precision. Recently, Tom\'e \cite{tome} explored some 
details of the joint evolution of two DKCA. She considered the problem of 
simultaneous updating of two replicas using pseudo-random
numbers; two pres\-crip\-tions were presented for the joint evolution:
one used by Martins {\it et al} \cite{martins} and another
introduced by Kohring and Schereckenberg \cite{alemao}.
Hinrichsen {\it et al} \cite{hyn} discovered a third phase in the DKCA diagram, 
using a thorough analysis of the damage spreading technique to split 
the active phase into three different regions: a chaotic region, where the 
damage spreads for every member of this family of dynamic procedures; an active 
region, where the damage heals for every member 
of this family, and another active region, where the 
damage spreads for a subset of the possible dynamic procedures and heals 
for the others. This new phase was obtained with a prescription
that updates the replicas using the minimal correlations.

In 1997, de Sales {\it et al} \cite{sales1} showed that the roughness 
exponent $\alpha$ can be used to classify the deterministic cellular 
automata (CA) described by Wolfram \cite{wolfram}. More recently \cite{sales2}, 
they showed that this exponent also can be used to 
detect the frozen/active transition in DKCA directly from the 
automata, without reference to order 
parameters or response functions. This method also can be used to detect phase
transitions in another kinds of models, such as the Potts model \cite{reid}.  
Beyond the roughness exponent $\alpha$,
the growth exponent $\beta$ is another critical exponent used to describe 
various roughening processes in the surface growth context 
\cite{barab,meakin}. 

In this work, we introduce the growth exponent method to identify phase
transitions. We apply this method to the one dimensional 
DKCA to detect the phase transitions directly from the automata and 
build the DKCA phase diagram. To obtain self-affine rough profiles, we use the 
accumulation method to perform a mapping to the solid-on-solid (SOS) model, 
and measure the time evolution of the roughness of the interface to obtain 
the exponent $\beta$. In the frozen/active transition, the exponent $\beta$ 
has a maximum, and two schemes are used to update the system \cite{nagy}: 
a symmetric scheme, that corresponds to a triangular lattice, and a 
nonsymmetric scheme. In the symmetric scheme we obtain
$\beta \sim 0.83$, consistent with the directed percolation (DP) prediction;
but in the nonsymmetric scheme, $\beta \sim 0.61$, an unusual value that was 
not expected. It is expected \cite{nagy}, rather, to find the same value 
for $\beta$ in the two schemes. 

Very recently, a quite similar method was used by 
Lauritsen and Alava \cite{lauri}, to study the Edwards-Wilkinson equation
with columnar noise, and by Dickman \cite{dickman}, to study the contact 
process. 

This studies do not consider damage spreading, and cannot detect 
the chaotic/non-chaotic transition.
To evidence this transition, we use the damage spreading ideas \cite{hans}: 
the difference between two replicas, which evolve with same dynamics, is used 
to perform the same SOS mapping as mentioned above, via the accumulation method. 
We use three different prescriptions for the simultaneous updating of
the replicas and obtain three different transition lines. 

In Section II, we introduce the growth exponent method and show the results 
obtained for the  
frozen/active transition. The damage spreading ideas and the chaotic/non-chaotic
transition are presented in the Section III. Finally, we let to the Section IV
our conclusions and acknowledgements.

\section{Frozen-Active Transition}
The DKCA consists of a linear chain of $L$ sites ($i=1,2,...,L$), with 
periodic boundary conditions, where each site has two possible states 
$\sigma = 0,1$ (frozen, active). 
The state of the system at time $t$ is given by the set \{$\sigma_{i}$\}. In 
contrast to the deterministic CA studied by Wolfram 
\cite{wolfram}, the DKCA is probabilistic, in the sense that the 
rules that update the system are given by conditional probabilities,
$P(\sigma_{i-1}(t), \sigma_{i+1}(t) | \sigma_{i}(t+1))$ (in the symmetric 
scheme). 
That is, the state of given site in time ($t+1$) depends in a 
probabilistic fashion upon the state of the 
two nearest neighbours at time $t$. In the DKCA, the conditional 
probabilities in the isotropic case are 
\[
P(0,1 | 1) = P(1,0 | 1) = p_1, ~~~P(1,1 | 1) = p_2,~~~P(0,0 | 1) = 0.
\]
Obviously, 
\[
P(\sigma_{i-1}, \sigma_{i+1} | 0) = 1 - P(\sigma_{i-1}, \sigma_{i+1} | 1)~.
\]
In the nonsymmetric scheme, the conditional probabilities are given by
$P(\sigma_{i-1}(t), \sigma_{i}(t) | \sigma_{i}(t+1))$, and 
$P(\sigma_{i-1}, \sigma_{i} | 0) = 1 - P(\sigma_{i-1}, \sigma_{i} | 1)~$.
The values of the conditional probabilities are the same in both schemes. 
In Figure 1 we reproduce a representation of these schemes.

Depending on the values of the parameters ($p_1,p_2$), the asymptotic 
($t \rightarrow \infty$) 
state of the system is either a {\it frozen} state, 
with all sites in state 0, or has a finite fraction of 
sites with value 1, the {\it active} state. This is a second order phase 
transition, characterized by universal critical exponents \cite{dk}. 

To study the phase diagram of the DKCA, we use the accumulation method, 
introduced
by Sales {\it et al.} \cite{sales2}, to obtain profiles in analogy
with solid-on-solid (SOS) models in ($1+1$) dimensions.  
This method consists in accumulating (or summing) all the values 
assumed by the variables $\sigma_{i}(t)$ during a 
given number $t$ of successive time steps
\beg\label{soma}
h_{i}(t) = \sum^{t}_{\tau=0} \sigma_{i}(\tau)~~.
\fim
 
The differences between the schemes become explicit at this point. In the
Figure 2, we can observe several profiles at criticality in both schemes, 
generated after we apply the accumulation method. It is evident that different
schemes lead to completely different profiles.
 
Thus, we obtain growth processes, the nature of whose correlations can be 
investigated through the analysis of the roughness $w(L,t)$ 
\cite{barab}. The roughness is defined by
\beg\label{rugos}
w^{2}(L,t) = \frac{1}{L} \sum^{L}_{i=1} \left( h_{i}(t) - \overline{h}(t) 
\right)^{2}~~,
\fim

\noindent where $\overline{h}(t)$ is the mean value of $h_{i}(t)$ at time $t$.
In fact, in order to consider the initial roughness of the profiles, we work 
with the fluctuation in roughness \cite{tales}
\beg\label{dw}
\delta w (L,t) = \sqrt{ w^{2}(L,t) - w^{2}(L,0) }~.
\fim
We expect that the behavior of $\delta w(L,t)$ has the form 
\beg\label{scal}
\delta w(L,t) \sim L^{\alpha} f \left( \frac{t}{L^{z}} \right)~~,
\fim

\noindent where $f(u)$ is a universal scaling function, $\alpha$ is the 
roughness exponent, $z = ( \alpha/ \beta)$ is the dynamic exponent and 
$\beta$ is the growth exponent. The function $f(u) = constant$ at large 
times ($t \gg L^{z}$) and $f(u) \sim u^{\beta}$ at short times ($t \ll L^{z}$).
So, at short times, we expect that $\delta w(L,t) \sim t^{\beta}$ and can 
measure $\beta$ calculating
the slope of the $\log - \log$ plot of $\delta w(L,t)$ versus $t$. 
The growth exponent denotes how the profile 
roughness grows with time: $\beta =1/2$ means that the profile is not 
correlated, and is analogous to that generated by random deposition 
\cite{barab}; if $\beta >1/2$, the profile tends to grow
more at the tips, which causes the roughness increases faster, in contrast to 
$\beta < 1/2$, where the 
valleys grows quickly and make the roughness increase more slowly. 

Typical results for the evolution of the roughness are showed 
in Figure 3. These results 
correspond to averages over 50 random initial configurations taken after 
$100000$ time steps in DKCA containing $L = 10000$ sites. Each curve in this 
graph (101 points) takes approximately one hour of CPU time on a Digital 
$500au$ workstation. 
We can observe that the roughness reaches the steady state in the frozen 
phase and grows indefinitely in the active phase. The values of $\beta$ 
at the transition are showed in Figure 4, for both schemes. The exponent 
$\beta$ is measured over 
more than three decades ($10 < t < 100000$  and $0.1 < \delta w < 1000$). 
Note that $\beta$ shows a maximum at the transition and tends to the value 
$\beta = 1/2$ quickly after the transition, remaining at this value
until $p_1 =1$. The value of the $\beta$ exponent at 
the transition depends of the scheme utilized in the DKCA. Nagy {\it et al} 
describe two schemes: a symmetric one, where the even (odd) sites are updated 
at even (odd) times; and a nonsymmetric one, where all sites are updated at 
each time step, but 
the neighbours of site $(i,t+1)$ are $(i-1,t)$ and $(i,t)$.
The symmetric scheme has been used in most previos studies of the model.
In the symmetric scheme, we find $\beta = 0.81(2)$, compatible with the 
universality class of directed percolation (DP) ($\beta \sim 0.84$) 
\cite{dickman}.
In the nonsymmetric scheme, we find $\beta = 0.61(2)$, clealy different from
the DP value. 

A finite size scaling analysis was made for the exponent $\beta$
and shows that the width of the peak vanishes when the 
size of system goes to infinity. The value of the exponent $\beta$ when 
the system size goes to infinity approaches $\beta \sim 0.83(2)$
at the frozen/active transition in the symmetric scheme, and is valid
for all values of $p_2 \neq 1$. 

On the line $p_2 = 1$, the system is mapped in the two-dimensional Ising model
\cite{kinzel},
and the $\beta$ value is significant greater in the two schemes:
$\beta = 0.99 (1)$ in the symmetric, and $\beta = 0.75(1)$ in the nonsymmetric.
The value in the symmetric scheme agrees with the literature \cite{hyn}, 
which predicts that at all points on phase boundary, except the line
$p_2=1$, are characterized by directed percolation (DP) exponents.
On the line $p_2=1$,
the model has been solved exactly, and can be mapped in the bi-dimensional
Ising model universality class, which leads to the value $\beta=1$ \cite{dk}.

The value $\beta > 1/2$ denotes the trend of the system 
to grow faster at the tips which can be 
understood as a preservation of active sites. At the transition, few 
sites remain active, which
causes $h_{i}(t)$ to grow only near active sites, contributing to an increase 
in the roughness. In the active phase, 
many sites are active, but uncorrelated, randomly increasing the height 
$h_{i}(t)$, so that $\beta = 1/2$.

The roughness exponent is defined only in the frozen phase because,
in the active phase, the steady state is not reached. We calculate
the Hurst exponent, $H$, very close to the transition and find, in 
the symmetric scheme, $H = 0.61(2)$. As noted by de 
Sales {\it et al} \cite{sales2} , there is a maximum in the $H$ exponent,
marking the phase transition. In the nonsymmetric scheme, the value of 
$H$ at the transition is much smaller, $\alpha \sim 0.25$, and the maximum
is not very clear.

\section{Active-Chaotic Transition}

Martins {\it et al} \cite{martins} used the damage spreading technique
to show that the active phase of DKCA can be split into two phases,
chaotic and non-chaotic. The order parameter of this transition
is the difference between two replicas with slightly
different initial configurations. They let the system evolve until it 
attains equilibrium, and then a replica of the automaton is created  
with some sites altered (damage). So the two replicas, one with states
$\sigma_{i}(t)$ the another with states $\varrho_{i}(t)$, evolve
with the same dynamics, and the difference between the automata  
\[
\Gamma_{i}(t) = | \sigma_{i}(t) - \varrho_{i}(t) |~.
\]
is measured.
The  fraction of sites in replica system that differ from their counterpart 
in the original system is called the Hamming distance, defined as 
\[
D_H (t) = \frac{1}{L}\sum_i \Gamma_i (t)~~.
\]
The stationary Hamming distance
is null in the non-chaotic phase and positive in the chaotic phase. 

To obtain the chaotic/non-chaotic boundary, we use a slightly different method, 
where the difference between the two automata is used to perform the same 
mapping to a SOS model as we did in the accumulation method
\beg\label{acc2}
h_{i}(t) = \sum^{t}_{\tau=0} \Gamma_{i}(\tau)~.
\fim
Thus, the profile generated by the difference of the two replicas 
behaves exactly as the profiles generated in the frozen/active boundary:
the roughness reaches a stationary value in the non-chaotic phase and grows 
indefinitely in the chaotic phase. This behavior
can be understood if we note that the difference between the replicas vanishes 
in the non-chaotic phase, which implies no contribution to the height 
$h_{i}(t)$, and is positive in chaotic phase, which implies in a 
persistent contribution to the height. The $\beta$ exponent again passes 
through a 
maximum at the chaotic/non-chaotic transition, and its value depends 
on the scheme utilized. 

Figure 5 shows $\beta$ for the chaotic/non-chaotic transitions, in
the symmetric scheme. To obtain this 
figure, we use an automaton with $L=10 000$ and let it evolve for $10^4$ 
time steps, at which time we create a replica
of the system with an ``initial damage'', by flipping a fraction of the 
sites ($\sim 10 \%$).
Then the replicas evolve with the same dynamics during the $10^5$ time steps, 
and the difference between them
is measured as a function of time. An average over $50$ realizations of 
initial damage was used.

We can locate the chaotic/non-chaotic phase transition, with the damage method,
either waiting or not waiting for
the original system to reach the steady state (first $10^4$ time steps). The 
only difference we note is that the value for the exponent $\beta$, 
at the frozen/active transition, is more precise and the cusp is more 
pronounced if we wait the original to reach the steady state. For the 
chaotic/non-chaotic transition we
obtain the same values with the two procedures, considering the statistical 
fluctuations.

At this point we have to emphasize the question of the dynamics of joint 
evolution of two CA's. Tom\'e \cite{tome} studied this joint 
evolution and showed that the CA's can evolve following two 
different prescriptions: prescription A, used by Martins {\it et al.}
\cite{martins}, which corresponds to updating both automata always using 
the same random number, and
prescription B, introduced by Kohring and Schereckenberg \cite{alemao}, 
which implies that one must sometimes use two different random numbers 
($z_1$ and $z_2$)
to update the original and the replica. The later occurs when we have 
$(\sigma_{i-1} + \sigma_{i+1} )=1$ and $(\varrho_{i-1} + \varrho_{i+1})=2$, or 
vice-versa.
We make simulations using these two prescriptions and verify that there are 
significant differences in the chaotic/non-chaotic boundary of the DKCA
$p_2-p_1$ phase 
diagram, as showed in Figure 6. These differences were discovered by 
Bagnoli \cite{bagnoli}, who studied the damage spreading
transition in the DKCA using a mean-field approximation; this work
confirms his prediction numerically. The
frozen/active transition line and the chaotic/non-chaotic transition line 
obtained with prescription A are very close to the the phase diagram 
found by Zebende and Penna \cite{zebende}.

In this phase diagram we also present a third prescription, C,
in which three random numbers, $z_1$, $z_2$ and $z_3$, are used to update the 
system. Defining 
\[U_i = \sigma_{i-1} + 2*\sigma_{i+1}\]    and   
\[V_i = \varrho_{i-1} + 2*\varrho_{i+1} ~~,\]  
we have the following cases: \newline
$\bullet$ if $U_i = V_i ~~\rightarrow$ we use the same random number ($z_1$) 
to update the original and the replica; \newline
$\bullet$ if $U_i = 1$ and $V_i = 2$ (or vice-versa) $\rightarrow$ we use
$z_1$ for the original and $z_2$ for the replica; \newline
$\bullet$ if $U_i = 1$ and $V_i = 3$ (or vice-versa) $\rightarrow$ 
we use $z_1$ for the original and $z_3$ for the replica. \newline
$\bullet$ if $U_i = 2$ and $V_i = 3$ (or vice-versa) $\rightarrow$ 
we use $z_2$ for the original and $z_3$ for the replica. \newline
To this prescription, the boundary is slighting above that of prescription B.

The difference between the phase boundaries appears to be due the different 
prescriptions for updating the systems in the damage spreading technique
employed to detect the chaotic/non-chaotic transition: prescription A 
corresponds to
maximal correlations between the random numbers; prescription B, to lower 
correlations, and prescription C to minimal correlations \cite{hyn,bagnoli}. 
We perform simulations using the third prescription in an attempt to reveal
the third phase transition, reported by Hinrichsen {\it et al}
\cite{hyn}, but are unable to detect that transition using prescription C, 
which is the prescription with minimal correlations.

\section{Conclusions}
In this work we propose a new method to obtain the phase diagram of the 
DKCA using the growth exponent $\beta$,
and detect the frozen/active boundary.
At the transition, the values of the exponent $\beta$ depend
on the scheme used to update the system: $\beta =0.83(2)$ for the
symmetric scheme, and $\beta =0.61(2)$ to the nonsymmetric scheme.
Next, we extend this method, using damage spreading, to 
obtain the chaotic/non-chaotic boundary. Finally, we study three different 
prescriptions for the joint evolution
of two DKCA's and construct the phase diagram, showed in Figure 6. 

The advantage of this method to determine the phase diagram of the DKCA is 
that we do not need to
wait for the system to reach the steady state, as in the methods used by 
Martins {\it et al} \cite{martins} and Zebende and Penna \cite{zebende},
thereby economizing computation time. In addition, the growth
exponent method can detect the chaotic/non-chaotic boundary much more 
clearly than the usual Hamming 
distance, which presents large fluctuations at the transition. This 
method can also be employed to detect phase transitions
in other models where the accumulation process can be used \cite {reid}. 
 
\noindent {\bf Acknowledgements}
The authors are indebted to M. L. Martins, T. J. P. Penna and R. Dickman
for fruitful discussions and suggestions. 
We also thank R. Dickman for helpful criticism of the manuscript. 
This work is supported by Brazilian agencies CNPq and Fapemig. 



\newpage

{\large \bf Figure captions}

\noindent {\bf Figure 1}
Schemes used to update the systems. 

\noindent {\bf Figure 2}
Evolution of profiles generated by the accumulation method. Above:
profiles generated in the symmetric scheme on a lattice with $L=1000$. 
Below: profiles generated in the nonsymmetric scheme, $L=1000$. Both
figures have the same initial profile and same sequence of random noise,
and the profiles were taken at the same instants of time in both schemes.
Each profile was taken after 1000 time steps. 

\noindent {\bf Figure 3} 
Evolution of the fluctuation in roughness $\delta w(L,t)$ with time $t$ 
in a log-log plot, for $p_2 = 0.95$ and five different values of $p_1$. 
We use $L=10000$ and $50$ samples. Note that the frozen/active transition
occurs when the roughness grows indefinitely ($p_1 = 0.595$ - 
filled symbols).

\noindent {\bf Figure 4} 
Evolution of growth exponent $\beta$ as a function of $p_1$, in the frozen/active
transition, in the two updating schemes; the system has $L=10 000$. Five 
different 
values of $p_2$ are shown. The maximum of $\beta$ indicates the transition. 
The symmetric scheme is represented by filled symbols and the nonsymmetric
scheme by open symbols. The lines are to guide the eye.

\noindent {\bf Figure 5}
Evolution of growth exponent $\beta$ as a function of $p_1$, using damage
spreading ideas to locate the chaotic/non-chaotic
transition. The system has $L=10000$. Four different values of $p_2$ are 
showed, for prescription A. 

\noindent {\bf Figure 6} 
DKCA $p_1-p_2$ phase diagram obtained via the growth exponent method. Note that the 
three prescriptions yield different boundaries for
the chaotic/non-chaotic transition. At $p_2 = 0$ we have $p_{1c} = 0.809(1)$ 
and at $p_1 = 1$ we have $p_{2c} = 0.3135(10)$, with all boundaries meeting at 
these points.

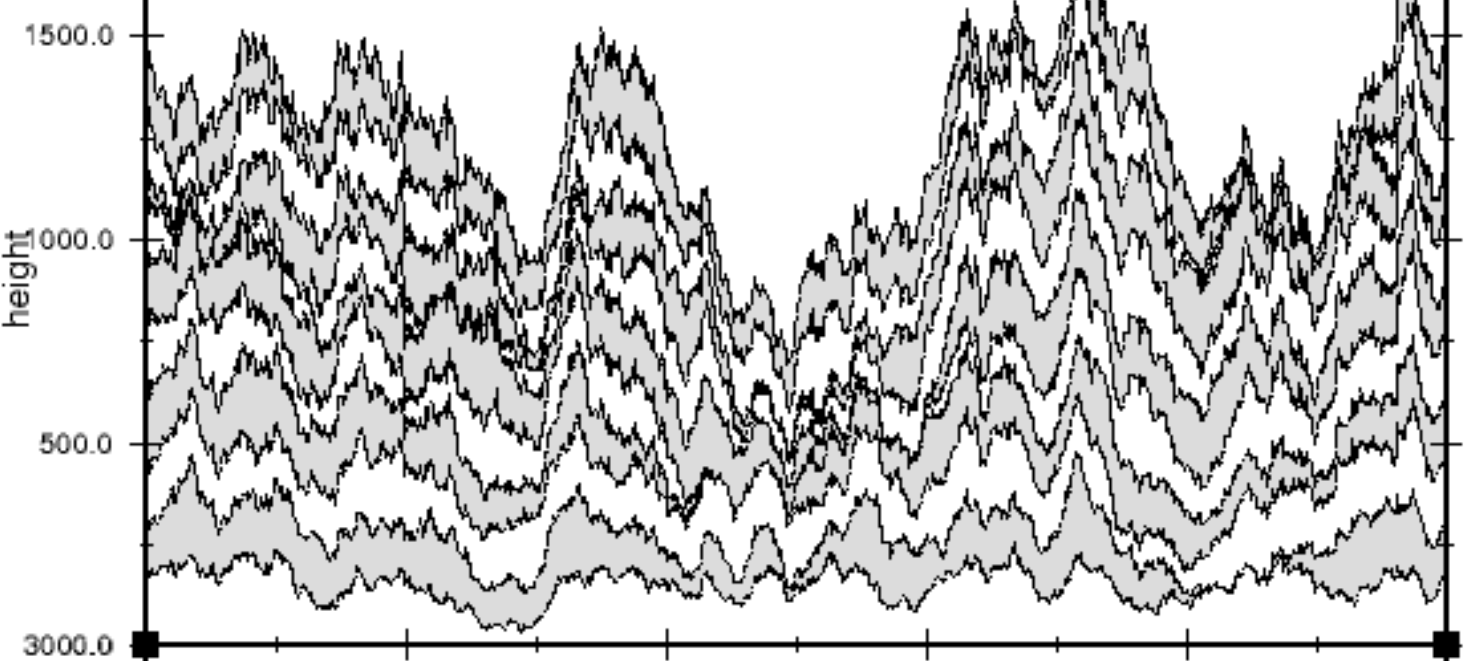
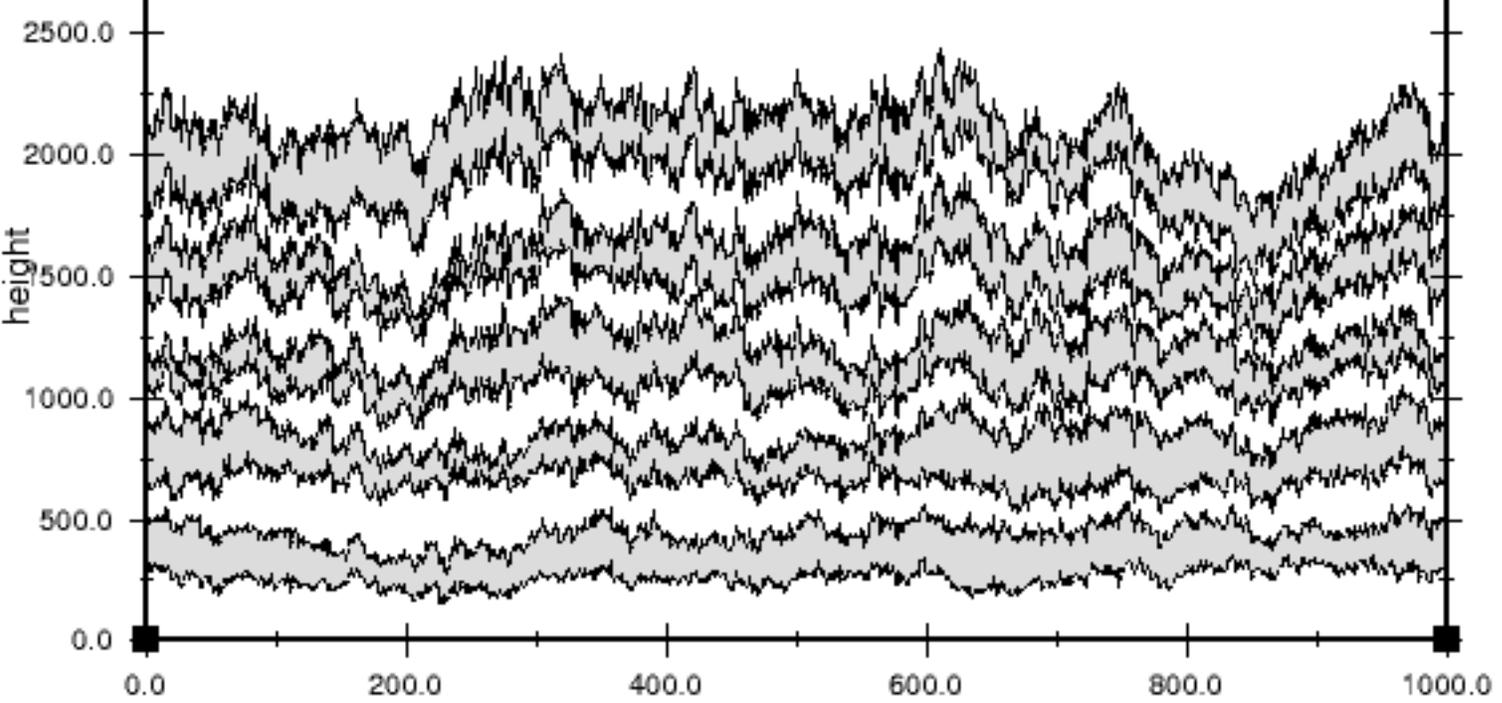
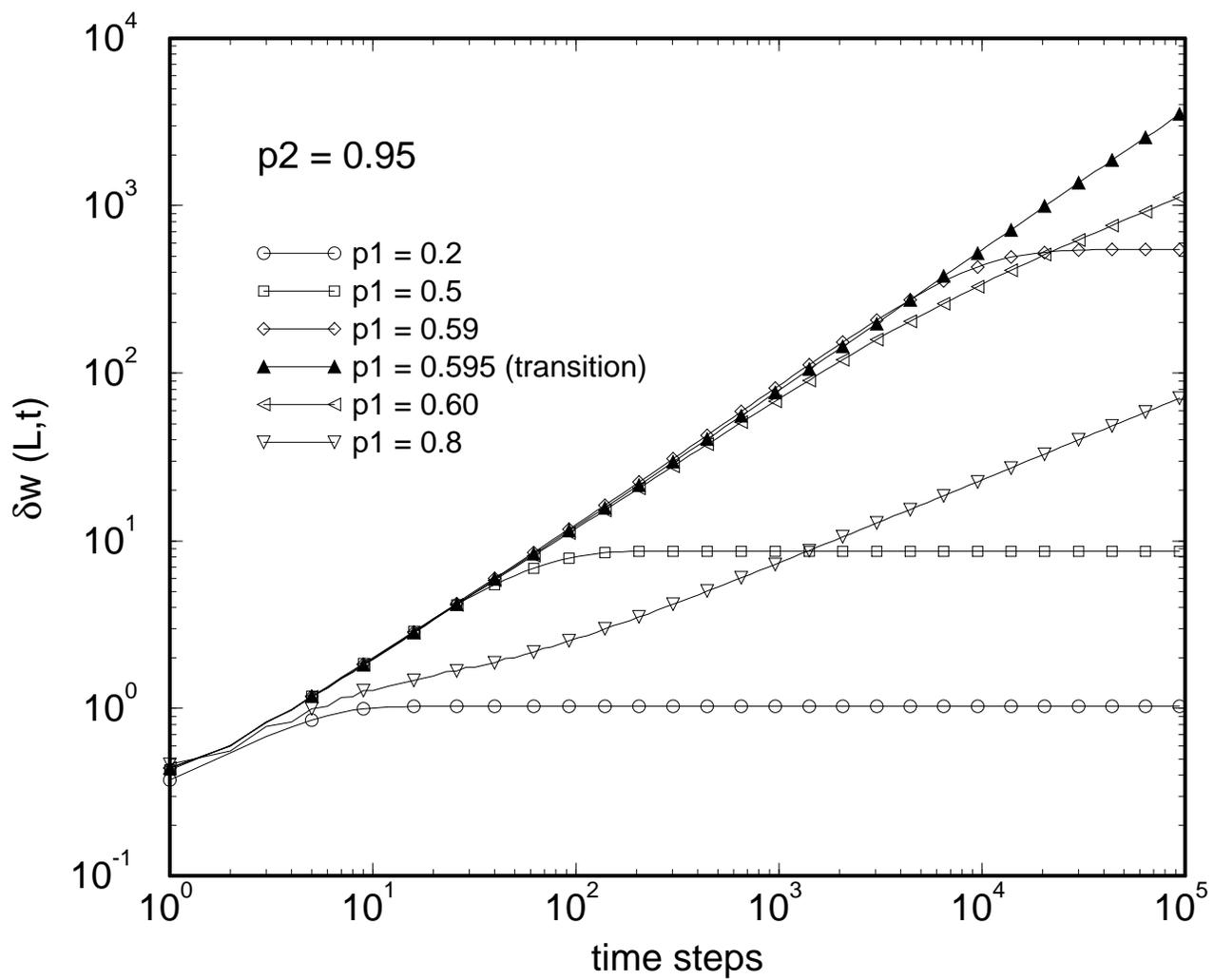
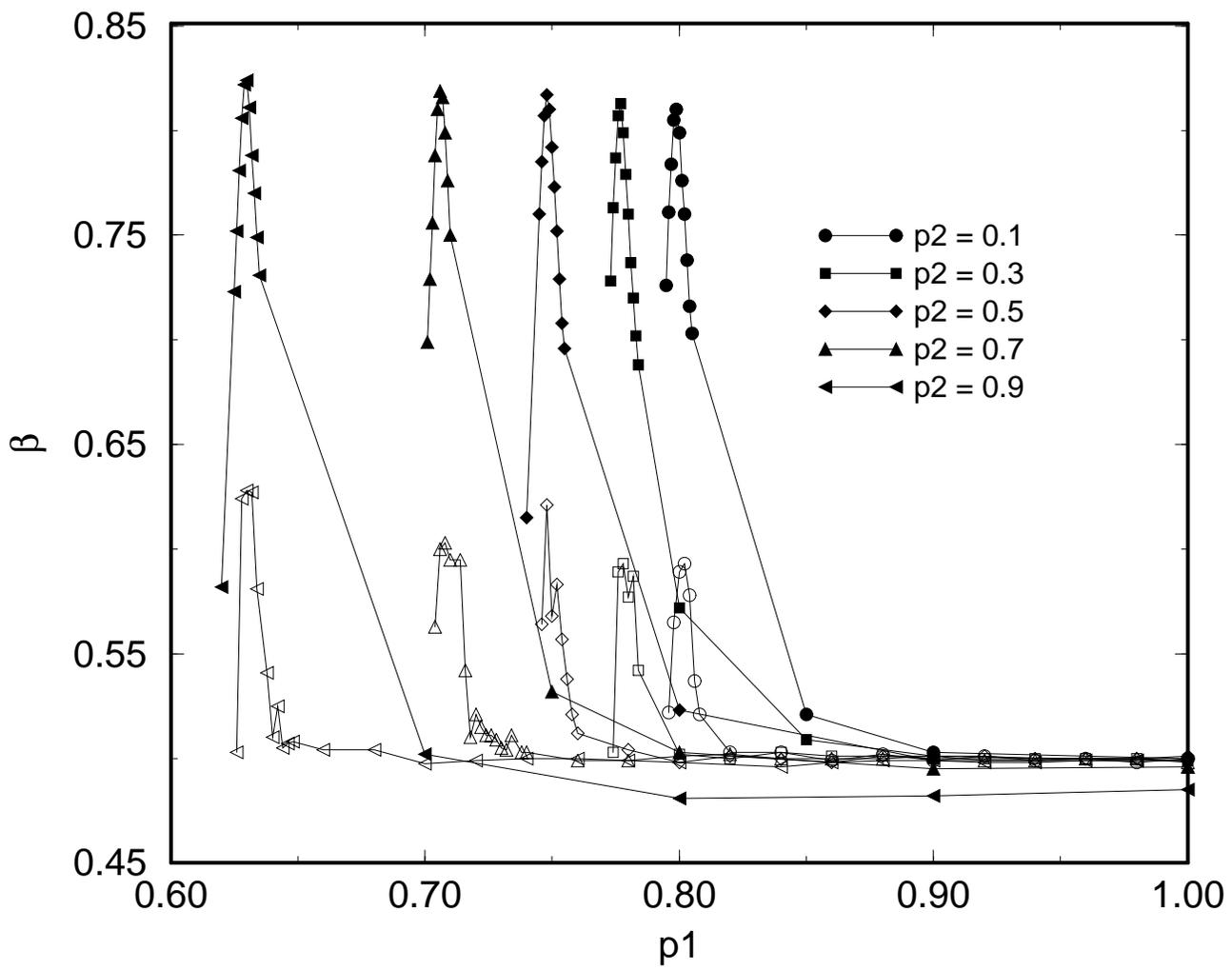
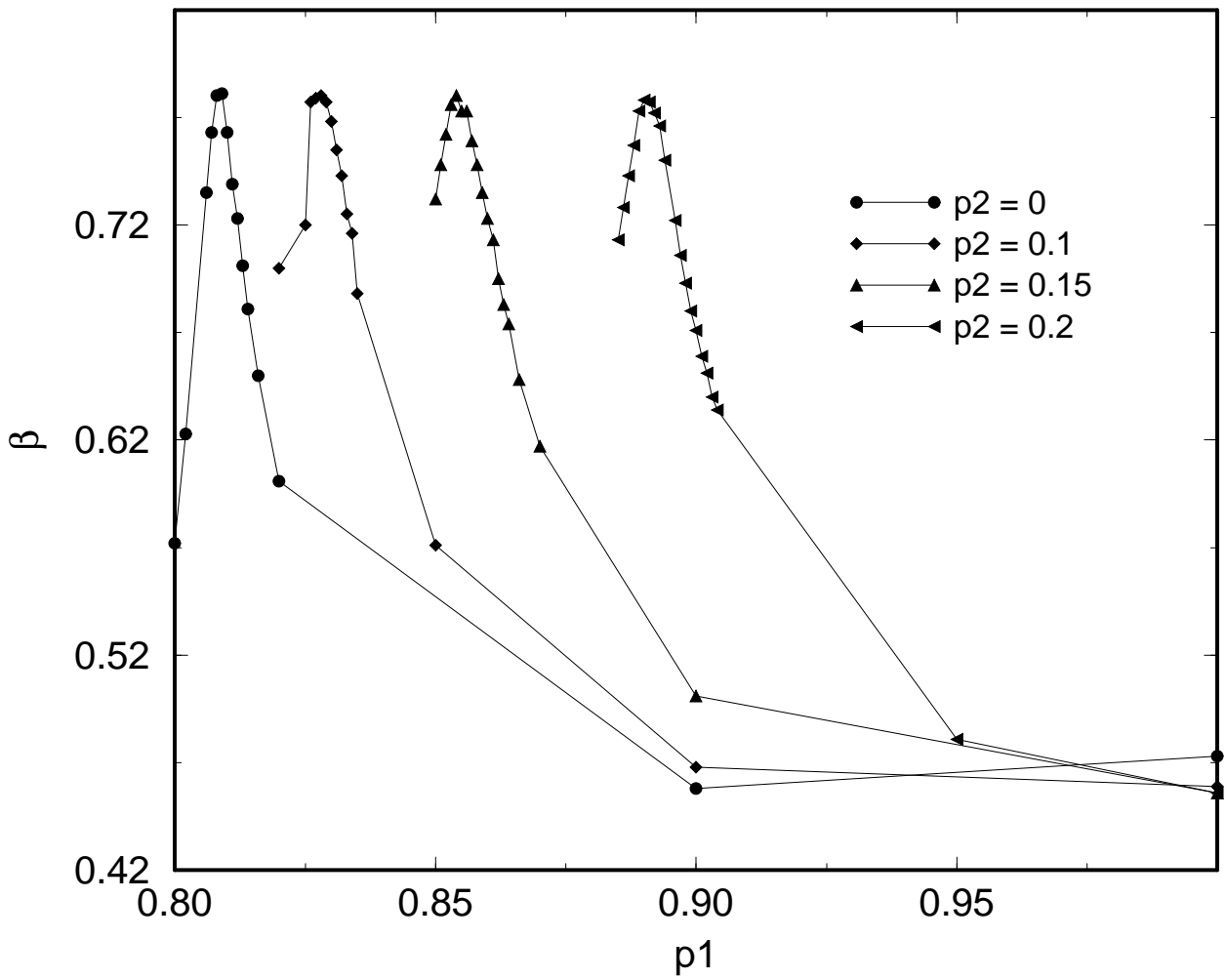
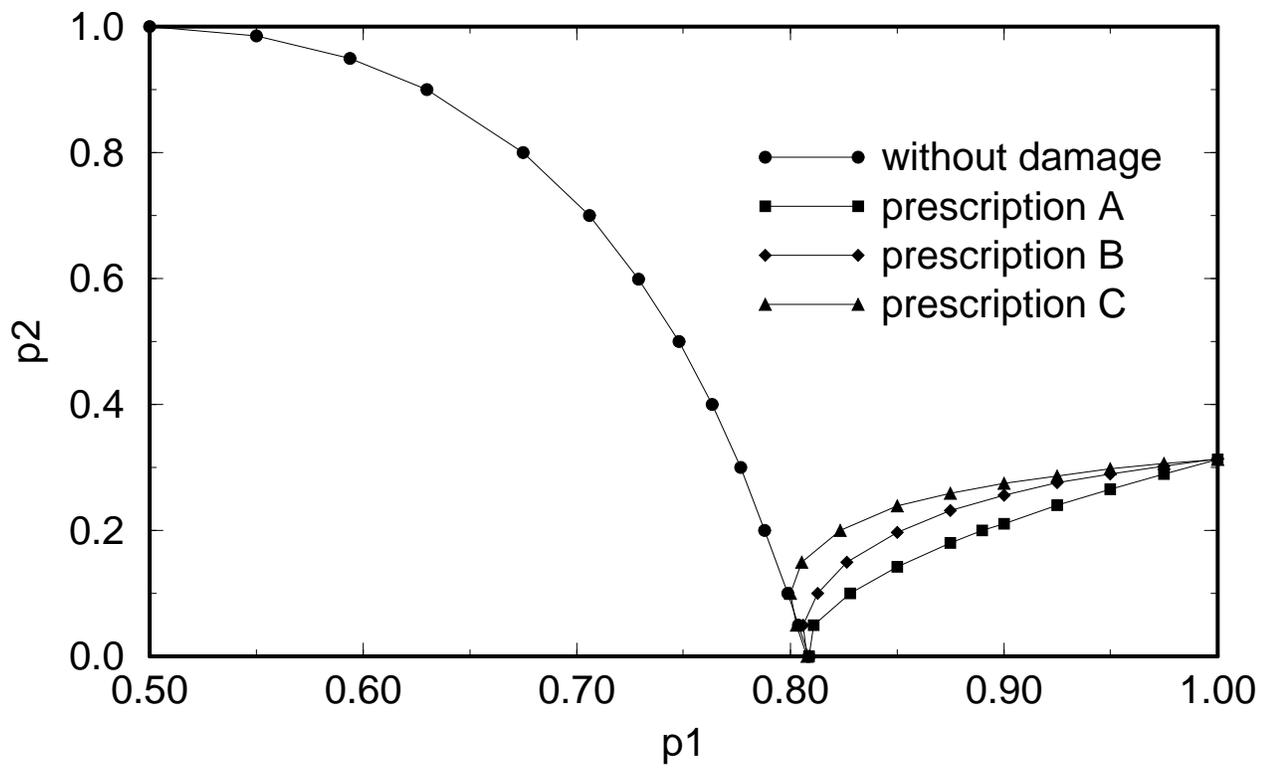
\end{document}